\documentstyle[12pt]{article}

\begin{document}

\title{The normalized energy eigenspinors of the Dirac field on  
anti-de Sitter spacetime}

\author{Ion I. Cot\u aescu\\ {\it The West University of Timi\c soara,}\\
{\it V. Parvan Ave. 4, RO-1900 Timi\c soara}}

\maketitle

\begin{abstract}
It is shown how  can be derived the normalized  energy eigenspinors 
of the free Dirac field on anti-de Sitter spacetime,  by using a  Cartesian 
tetrad gauge where the separation of spherical variables  can be done like in 
special relativity. 

Pacs: 04.62.+v
\end{abstract}
\

A  fundamental but difficult problem in general relativity is to find the 
analytical forms of the quantum free fields in given local charts. The 
difficulties arise because  the particular solutions of the field equations 
that describe the one-particle quantum modes are strongly dependent on the 
procedure of separation of variables and, implicitly, on the choice of the 
holonomic coordinates.  For the Dirac field the situation is more complicated  
since  the form of the field equations depends, in addition, on the tetrad 
gauge in which one works. In these conditions it is helpful to exploit the 
effects of the global symmetries of background that guarantees the existence of 
conserved observables.  

An important case is that of the Dirac equation on spacetimes with  spherically 
symmetric (central) static charts that have the global symmetry of the group 
$T(1)\otimes SO(3)$, of time translations and  rotations of the Cartesian space 
coordinates. For these backgrounds we  have proposed recently a Cartesian 
tetrad gauge \cite{C0,C} which assures the covariance under this group of the 
Dirac equation in  Cartesian  coordinates, the whole theory acquiring thus the 
global symmetry of the background. Consequently,  the energy and  angular 
momentum are conserved  like in special relativity from which we can take over 
the method of separation of variables in spherical coordinates. 
In fact, our gauge defines Cartesian local (unholonomic) frames which play 
the same role as the Cartesian natural frame of the Minkowski spacetime, since 
their axes are just those of projections of the orbital  angular momentum. This 
allowed us to separate the spherical variables in terms of usual angular 
spinors such that  all the constants involved in  separation of variables get  
physical meaning as eigenvalues of the familiar  set of commuting operators 
$H$, $J^{2}$, $J_{3}$ and $K$ \cite{BJD,TH}. On this way we have found a 
complete formulation of the radial problem of the Dirac equation on central 
static backgrounds, deriving the radial equations and the form of the radial 
scalar product in the general case \cite{C0,C}. Similar results have been 
obtained in Ref.\cite{F} by using a different method based on an extended 
gauge. 

In this approach  the Dirac equation on  anti-de Sitter (AdS) or de-Sitter 
 static backgrounds can be analytically solved in terms of Gauss 
hypergeometric functions. Thus we have obtained the  energy spectra and the 
 energy eigenspinors up to normalization factors, in both these cases 
\cite{C,C1}. However, the normalization of the energy eigenspinors is very 
important for the further developments of the quantum theory. In the de Sitter 
spacetime, where the energy spectrum is continuous, it is less probable to find 
an efficient normalization procedure but on AdS  the energy spectrum is 
discrete and, therefore, the normalization may be done in usual way. This is 
just the problem we would like to discuss here.    
Our aim is to present how can be selected the quantum modes and to write down 
the normalized energy eigenspinors of the regular modes of the Dirac field on 
AdS, in spherical 
coordinates (and natural units with $\hbar=c=1$). Moreover, we establish their 
orthogonality properties with respect to the relativistic scalar product. 
In order to introduce the framework we need, we start with a brief review of 
some general results concerning the radial Dirac problem in our gauge and then 
turn to the AdS case.

Let us denote by $\psi$ the free Dirac field of mass $M$ defined on 
a  central static chart with  Cartesian coordinates $(t,{\bf x})$ or 
spherical natural coordinates $(t, r, \theta, \phi)$, commonly related to 
the Cartesian ones. The space domain of this chart is $D=D_{r}\times S^{2}$,
i.e. $r\in D_{r}$ while $\theta$ and $\phi$ cover the sphere $S^{2}$. In the 
following it is useful to consider only the charts where the radial coordinate 
$r=|{\bf x}|$ is defined such that $g_{rr}=-g_{00}$. Then the tetrad fields in 
our Cartesian gauge \cite{C0,C} depend on two  arbitrary functions of $r$, 
denoted by  $v$ and $w$, which give the line element 
\begin{equation}\label{(muvw)}
ds^{2}=w^{2}\left[dt^{2}-dr^{2}-
\frac{r^2}{v^2}(d\theta^{2}+\sin^{2}\theta d\phi^{2})\right].
\end{equation}
and determine a suitable form of the Dirac equation. This has  particular 
positive frequency solutions of energy $E$,  
\begin{equation}\label{(spin)}
\psi^{(+)}_{E,j,m_{j},\kappa_{j}}(t,{\bf x})=
U_{E,j,m_{j},\kappa_{j}}({\bf x})e^{-iEt}\,, 
\end{equation}
where
\begin{eqnarray}
U_{E,j,m_{j},\kappa_{j}}({\bf x})&=&
U_{E,j,m_{j},\kappa_{j}}(r,\theta,\phi)\label{(u)}\\
&=&\frac{v(r)}{rw(r)^{3/2}}[f^{+}(r)\Phi^{+}_{m_{j},\kappa_{j}}(\theta,\phi)
+f^{-}(r)\Phi^{-}_{m_{j},\kappa_{j}}(\theta,\phi)]\nonumber
\end{eqnarray}
are the particle-like energy eigenspinors expressed in terms of radial wave 
functions $f^{\pm}$ and  four-component angular spinors $\Phi^{\pm}_{m_{j}, 
\kappa_{j}}$, known from special relativity \cite{TH}. We remind that they  
are orthogonal to each other being completely determined by the angular 
quantum numbers, $j$ and $m_{j}$, and 
the value of $\kappa_{j}=\pm (j+1/2)$ \cite{TH,BJD}. Moreover, they are 
normalized to unity with respect to their own angular scalar product. Thus 
the problem of the angular motion is solved in the same manner as in 
the flat spacetime. Obviously, the radial problem is different.    
The radial functions $f^{\pm}$ are  solutions of a pair of radial equations 
\cite{C0,C} that can be written in compact form as the eigenvalue problem 
\begin{equation}
H{\cal F}=E{\cal F}
\end{equation}
of the radial Hamiltonian 
\begin{equation}
H=\begin{array}{|cc|}
    Mw& -\frac{\textstyle d}{\textstyle dr}+\kappa_{j}\frac{\textstyle v}
{\textstyle r}\\
&\\
  \frac{\textstyle d}{\textstyle dr}+\kappa_{j}\frac{\textstyle v}
{\textstyle r}& -Mw
\end{array}\,,
\end{equation}
in the space of  two-component vectors,  ${\cal F}=|f^{+}, f^{-}|^{T}$, 
where the radial scalar product is \cite{C0,C} 
\begin{equation}\label{(spf)}
({\cal F}_{1},{\cal F}_{2})=\int_{D_{r}}dr\, 
{\cal F}_{1}^{\dagger}{\cal F}_{2}\,.
\end{equation}
This selects the "good" radial wave functions (i.e. square integrable 
functions or tempered distributions) which enter in the structure of the 
particle-like energy eigenspinors  (\ref{(u)}).  We note that the  
antiparticle-like energy eigenspinors  can be obtained directly by using the 
charge conjugation \cite{C}.

In the case of AdS spacetime it is convenient to consider the  
central static chart with the line element  \cite{AIS}
\begin{equation}\label{(le)}
ds^{2}=\sec^{2}\omega r \left[dt^{2}-dr^{2}-\frac{1}{\omega^{2}}
\sin^{2}\omega r~ (d\theta^{2}+\sin^{2}\theta~d\phi^{2})\right].
\end{equation} 
The radial domain of this chart is $D_{r}=[0,\pi/2\omega)$ because of the event 
horizon at $r=\pi/2\omega$. We specify that here we take 
$t\in (-\infty, \infty)$ which defines in fact the universal covering spacetime 
(CAdS) of AdS \cite{AIS}. Now, from (\ref{(le)}) we can 
identify the functions  
$w(r)=\sec \omega r$ and $v(r)=\omega r \csc \omega r$.
With their help and by using the notation $k=M/\omega$ (i.e. 
$Mc^{2}/\hbar\omega$ in usual units), we obtain  
\begin{equation}
H=\begin{array}{|cc|}
    \omega k\sec \omega r& -\frac{\textstyle d}{\textstyle dr}+
\omega\kappa_{j}\csc\omega r\\
&\\
       \frac{\textstyle d}{\textstyle dr}+\omega\kappa_{j}\csc\omega r
& -\omega k\sec \omega r
\end{array}\,.
\end{equation}
This Hamiltonian  has a hidden supersymmetry that can be pointed out 
with the help of the local rotation  ${\cal F}\to 
\hat{\cal F}=R{\cal F}=|\hat f^{+},\hat f^{-}|^{T}$
produced by   
\begin{equation}\label{(uder)} 
R(r)=\begin{array}{|cc|}
    \cos \frac{\textstyle \omega r}{\textstyle 2}&-\sin 
\frac{\textstyle \omega r}{\textstyle 2}\\
&\\    
\sin \frac{\textstyle \omega r}{\textstyle 2}&\cos 
\frac{\textstyle \omega r}{\textstyle 2}
\end{array}\,.
\end{equation}
Indeed, after a few manipulation we find that the 
transformed (i.e. rotated and translated) Hamiltonian,   
\begin{equation}\label{(newh)}
\hat H =RHR^{T}-\frac{\omega}{2} 1_{2\times 2},
\end{equation}
has supersymmetry since it  has  the requested specific form  
with diagonal constant terms \cite{TH} and a P\" oschl-Teller-like \cite{PT}
superpotential \cite{C}.

In these conditions the new eigenvalue problem
\begin{equation}\label{(trrp)}
\hat H \hat{\cal F}=\left(E-\frac{\omega}{2}\right)\hat{\cal F},
\end{equation} 
involving the transformed radial wave functions $\hat f^{\pm}$,   
leads to a pair of  second order equations giving \cite{C}
\begin{eqnarray}
\hat f^{\pm}(r)&=&N_{\pm}
\sin^{2s_{\pm}}\omega r\cos^{2p_{\pm}}\omega r \label{(gsol)}\\
&&\times F\left(s_{\pm}+p_{\pm}-\frac{\epsilon}{2},
s_{\pm}+p_{\pm}+\frac{\epsilon}{2}, 2s_{\pm}+\frac{1}{2}, \sin^{2}\omega r
\right).\nonumber
\end{eqnarray}
where  $F$ are the Gauss hypergeometric functions \cite{AS}. Their 
real parameters are defined as, $\epsilon=E/\omega-1/2$ and     
\begin{eqnarray}
2s_{\pm}(2s_{\pm}-1)&=&\kappa_{j}(\kappa_{j}\pm 1)\,,\label{(2s)}\\ 
2p_{\pm}(2p_{\pm}-1)&=&k(k\mp 1)\,,\label{(2p)} 
\end{eqnarray}
while $N_{\pm}$ are normalization factors. The next step is to select the 
suitable values 
of these parameters  and to calculate $N_{+}/N_{-}$ such that the functions 
$\hat f^{\pm}$ should be solutions of the transformed radial problem 
(\ref{(trrp)}), with  a good physical meaning. This can be achieved only 
when $F$ is a polynomial selected by a suitable quantization condition 
since otherwise $F$ is strongly divergent for $\sin^{2}\omega r\to 1$. 
Then the  functions $\hat f^{\pm}$ will be square integrable 
with  normalization factors calculated according to the condition
\begin{equation}\label{(norm)}
({\cal F},{\cal F})= 
 (\hat{\cal F},\hat{\cal F})
=\int_{D_{r}}dr\,\left( |\hat f^{+}(r)|^{2}+
|\hat f^{-}(r)|^{2}\right)=1\,, 
\end{equation}  
resulted from the fact that the matrix (\ref{(uder)}) is orthogonal.

The discrete energy spectrum is given by the  particle-like CAdS quantization 
conditions
\begin{equation}\label{(quant)}
\epsilon=2 (n_{\pm}+s_{\pm}+p_{\pm})\,, \quad \epsilon>0\,,
\end{equation}
that must be compatible with each other, i.e.
\begin{equation}\label{(comp)}
n_{+}+s_{+}+p_{+}=n_{-}+s_{-}+p_{-}.
\end{equation}
Hereby we see  that there is only one independent {\em radial} 
quantum number, $n_{r}=0,1,2,...$. 
In addition, we shall use  the orbital quantum number $l$ of  
the spinor $\Phi^{+}_{m_{j},\kappa_{j}}$ \cite{BJD}, as an auxiliary quantum 
number. 
On the other hand, if we express  (\ref{(gsol)}) in terms of Jacobi 
polynomials, we observe that these functions remain square integrable for 
$2s_{\pm}>-1/2$ and $2p_{\pm}>-1/2$. Since $l=0,1,2...$  
we are forced to select only  the positive solutions of 
Eqs.(\ref{(2s)}). The different solutions of Eqs.(\ref{(2p)}) defines 
the boundary conditions of the allowed quantum modes, like in the case of 
the scalar modes \cite{BF}. We say that for 
$k>-1/2$ the values $2p_{+}=k$ and $2p_{-}=k+1$ define the boundary 
conditions of {\em regular} modes. The other possible values,     
$2p_{+}=-k+1$ and $2p_{-}=-k$, define the {\em irregular} modes when 
$k<1/2$. Obviously, for $-1/2<k<1/2$ both these modes are possible. 
We note that the AdS quantization conditions require, in addition, $k$ to be a 
half integer. Then it is clear that the domains of $k$ corresponding to the 
regular and respectively irregular  
modes can not overlap with each other.  Anyway, in our opinion, the 
problem of the meaning of the irregular modes as well as that of the 
relation between these kind of modes is sensitive and may be carefully 
analyzed. For this reason we restrict ourselves 
to write down only the energy eigenspinors of the regular modes on CAdS.   

Let us  take first $\kappa_{j}=-(j+1/2)=-l-1$. Then the positive solutions of 
(\ref{(2s)}) are $2s_{+}=l+1$ and $2s_{-}=l+2$ 
while, according to (\ref{(comp)}), we must have 
$n_{+}=n_{r}$ and $n_{-}=n_{r}-1$.
For these values of  parameters, the functions $\hat f^{\pm}$ given by 
(\ref{(gsol)}) and (\ref{(quant)}) represent a 
solution of the transformed radial problem (\ref{(trrp)}) if and only if
\begin{equation}\label{(npnm)}
\frac{N_{-}}{N_{+}}=-\frac{2n_{r}}{2l+3}.
\end{equation} 
Furthermore, it is easy to express (\ref{(gsol)}) in terms of 
Jacobi polynomials and to calculate the normalization factors from  
(\ref{(norm)}).   
Thus we arrive at the result,  
\begin{eqnarray}
\hat f^{+}(r)_{|\kappa_{j}=-(j+1/2)}&=& 
N\left[\frac{n_{r}+k+l+1}{n_{r}+l+\frac{1}{2}}\right]^{\frac{1}{2}}\nonumber\\
&&\times \sin^{l+1}\omega r \cos^{k}\omega r 
P_{n_{r}}^{(l+\frac{1}{2},k-\frac{1}{2})}(\cos 2\omega r)\,,
\label{(1)}\\
\hat f^{-}(r)_{|\kappa_{j}=-(j+1/2)}&=&-N \left[\frac{n_{r}+k+l+1}{n_{r}+l+
\frac{1}{2}}\right]^{\frac{1}{2}}\nonumber\\
&&\times \sin^{l+2}\omega r \cos^{k+1}\omega r 
P_{n_{r}-1}^{(l+\frac{3}{2},k+\frac{1}{2})}(\cos 2\omega r)\,,
\nonumber
\end{eqnarray}
where
\begin{equation}
N=\eta\sqrt{2\omega}\left[\frac{n_{r}!\,\Gamma(n_{r}+k+l+1)}
{\Gamma(n_{r}+l+\frac{1}{2})\Gamma(n_{r}+k+\frac{1}{2})}\right]^{\frac{1}{2}}
\,.
\end{equation} 
is defined up to the phase factor $\eta$. Notice that from (\ref{(npnm)}) we 
understand that the second equation of (\ref{(1)}) gives $\hat f^{-}=0$ for 
$n_{r}=0$.

For $\kappa_{j}=j+1/2=l$ we use the same procedure finding that 
$2s_{+}=l+1$, $2s_{-}=l$, $n_{+}=n_{-}=n_{r}$ and
\begin{equation}
\frac{N_{-}}{N_{+}}=\frac{2l+1}{2n_{r}+2k+1}\,.
\end{equation} 
In this case the normalized radial wave functions are 
\begin{eqnarray}
\hat f^{+}(r)_{|\kappa_{j}=j+1/2}&=&
N\left[\frac{n_{r}+k+\frac{1}{2}}{n_{r}+l+\frac{1}{2}}\right]^{\frac{1}{2}}
\nonumber\\
&&\times \sin^{l+1}\omega r \cos^{k}\omega r 
P_{n_{r}}^{(l+\frac{1}{2},k-\frac{1}{2})}(\cos 2\omega r)\,,
\label{(2)}\\
\hat f^{-}(r)_{|\kappa_{j}=j+1/2}&=&
N \left[\frac{n_{r}+l+\frac{1}{2}}{n_{r}+k+\frac{1}{2}}\right]^{\frac{1}{2}}
\nonumber\\
&&\times\sin^{l}\omega r \cos^{k+1}\omega r 
P_{n_{r}}^{(l-\frac{1}{2},k+\frac{1}{2})}(\cos 2\omega r)\,.
\nonumber
\end{eqnarray}

The energy levels result from (\ref{(quant)}). Bearing in mind that $\omega k=
M$ and $\omega\epsilon=E-\omega/2$, and by using the {\em  main} quantum number 
$n=2n_{r}+l$ we obtain \cite{C} 
\begin{equation}\label{(enlev)}
E_{n}=M+\omega\left(n+\frac{3}{2}\right)\,,\quad n=0,1,2,....
\end{equation} 
These levels are degenerated. For a given $n$ our auxiliary quantum number $l$ 
takes either all the odd values from $1$ to $n$, if $n$ is odd, or the even 
values from $0$ to $n$, if $n$ is even. In both cases  we have $j=l\pm 1/2$ 
for each $l$, which means that $j=1/2,3/2,...,n+1/2$. The selection rule for 
$\kappa_{j}$ is more complicated since it is determined by both the 
quantum numbers  $n$ and $j$. 
If $n$ is even then the even $\kappa_{j}$ are positive while the odd 
$\kappa_{j}$ are negative. For odd $n$ we are in the opposite situation, with 
odd positive or even negative values of $\kappa_{j}$. Thus it is clear that 
for each given pair $(n,j)$ we have only one value of $\kappa_{j}$. With these 
specifications and by taking into account that for each $j$ we have $2j+1$ 
different values of $m_{j}$, we can conclude that the degree of degeneracy of 
the level $E_{n}$ is $(n+1)(n+2)$. 

The solutions (\ref{(1)}) and (\ref{(2)}) are completely determined by 
the values of $n$ and $j$ since, according to the above rules, we have  
\begin{equation}
2n_{r}=n-j-\frac{1}{2}(-1)^{n+j+1/2}\,,\quad
l= j+\frac{1}{2}(-1)^{n+j+1/2}\,.
\end{equation}
For this reason, we  denote by $\hat f^{\pm}_{n,j}$ the radial wave functions 
(\ref{(1)}) and (\ref{(2)}). With their help we can write  
the functions $f^{\pm}_{n,j}$ (i.e. the components of ${\cal F}$)  by 
using the inverse of (\ref{(uder)}). Then from (\ref{(u)}) we find the 
definitive form of the normalized particle-like energy eigenspinors of the 
regular modes,   
\begin{eqnarray}\label{(defu)}
U_{n,j,m_{j}}({\bf x})&=&\omega \csc \omega r \cos^{3/2}\omega r\nonumber\\ 
&&\times\left[
\left(\cos\frac{\omega r}{2}\hat f^{+}_{n,j}(r)+\sin\frac{\omega r}{2}
\hat f^{-}_{n,j}(r)\right)\Phi^{+}_{m_{j},\kappa_{j}}(\theta,\phi)\right.\\   
&&+\left.\left(-\sin\frac{\omega r}{2}\hat f^{+}_{n,j}(r)+
\cos\frac{\omega r}{2}
\hat f^{-}_{n,j}(r)\right)\Phi^{-}_{m_{j},\kappa_{j}}(\theta,\phi)\right].
\nonumber
\end{eqnarray}
The antiparticle-like energy eigenspinors can be derived directly by using the 
charge conjugation \cite{BJD}. These are  
\begin{equation}
V_{n,j,m_{j}}=(U_{n,j,m_{j}})^{c}\equiv C (\overline{U}_{n,j,m_{j}})^{T} 
\,,\quad C=i\gamma^{2}\gamma^{0}\,. 
\end{equation}
Furthermore, we can verify that all these normalized energy eigenspinors 
have good orthogonality properties obeying
\begin{eqnarray}
&&\int_{D}d^{3}x\,\mu({\bf x})
\overline{U}_{n,j,m_{j}}({\bf x}) \gamma^{0} U_{n',j',m_{j}'}({\bf x})\\
&&~~~~~~~~~~~~~=\int_{D}d^{3}x\,\mu({\bf x})
\overline{V}_{n,j,m_{j}}({\bf x}) \gamma^{0} V_{n',j',m_{j}'}({\bf x})
=\delta_{n,n'}\delta_{j,j'}\delta_{m_{j},m_{j}'}\,,\nonumber\\
&&\int_{D}d^{3}x\,\mu({\bf x})
\overline{U}_{n,j,m_{j}}({\bf x}) \gamma^{0} V_{n',j',m_{j}'}({\bf x})\\
&&~~~~~~~~~~~~~=\int_{D}d^{3}x\,\mu({\bf x})
\overline{V}_{n,j,m_{j}}({\bf x}) \gamma^{0} U_{n',j',m_{j}'}({\bf x})=0\,,
\nonumber
\end{eqnarray}
where  
\begin{equation}\label{(mu)}
\mu({\bf x})=\frac{w(r)^3}{v(r)^2}=\frac{1}{\omega^{2} r^2}
\sin^{2}\omega r\, {\rm sec}^{3}\omega r 
\end{equation}
is the specific relativistic weight function \cite{C}. Of course, the 
factors $\omega$ and $1/\omega^2$ can be removed simultaneously from  
(\ref{(defu)}) and respectively (\ref{(mu)}).

The final result is that for $M\ge\omega/2$, when only regular modes are allowed, 
the quantum Dirac field on CAdS reads 
\begin{equation}
\psi(t,{\bf x})=\sum_{n,j,m_{j}}\left[U_{n,j,m_{j}}({\bf x})e^{-iE_{n}t}
a_{n,j,m_{j}}+V_{n,j,m_{j}}({\bf x})e^{iE_{n}t}b^{\dagger}_{n,j,m_{j}}
\right]\,.
\end{equation}
Our preliminary calculations indicate that the particle ($a$, $a^{\dagger}$) 
and antiparticle ($b$, $b^{\dagger}$) operators must  satisfy usual  
anticommutation relations from which the non-vanishing ones are   
\begin{equation}
\{a_{n,j,m_{j}}, a_{n',j',m_{j}'}^{\dagger}\}=
\{b_{n,j,m_{j}}, b_{n',j',m_{j}'}^{\dagger}\}
=\delta_{n,n'}\delta_{j,j'}
\delta_{m_{j},m_{j}'}.
\end{equation}
The argument is that then the one-particle operators ($Q$, $H$, etc.) derived 
from the Noether theorem  have similar structures and properties like 
those of the usual quantum field theory in flat spacetime. Hence we may have 
all the basic elements we need to construct the propagation theory of the 
regular modes of the Dirac field on CAdS, as the first step to a future theory 
of interacting quantum fields on this background.

\end{document}